\documentclass[preprint,a4paper]{aastex63}

\usepackage{graphicx}
\usepackage{bm}
\usepackage{amsmath,amssymb}
\usepackage{natbib}
\usepackage{mathtools}
\usepackage{epsf}
\usepackage[figuresright]{rotating}
\usepackage{mathrsfs}

\received{April 30, 2021}
\revised{January 30, 2022}
\accepted{February 14, 2022}

\shorttitle{Comparison of Two Methods for Calculating Magnetic Helicity}
\shortauthors{Wang et al.}

\begin{document}

\title{Comparison of Two Methods for Calculating Magnetic Helicity in the Solar Corona}

\correspondingauthor{Quan Wang}
\email{wangquan@nao.cas.cn}

\author[0000-0003-3142-217X]{Quan Wang}
\affiliation{Key Laboratory of Solar Activity, National Astronomical Observatories, Chinese Academy of Sciences, 100101 Beijing, People's Republic of China}
\affiliation{School of Astronomy and Space Sciences, University of Chinese Academy of Sciences, 100049 Beijing, People's Republic of China}

\author[0000-0002-2967-4522]{Shangbin Yang}
\affiliation{Key Laboratory of Solar Activity, National Astronomical Observatories, Chinese Academy of Sciences, 100101 Beijing, People's Republic of China}
\affiliation{School of Astronomy and Space Sciences, University of Chinese Academy of Sciences, 100049 Beijing, People's Republic of China}

\author[0000-0002-3141-747X]{Mei Zhang}
\affiliation{Key Laboratory of Solar Activity, National Astronomical Observatories, Chinese Academy of Sciences, 100101 Beijing, People's Republic of China}
\affiliation{School of Astronomy and Space Sciences, University of Chinese Academy of Sciences, 100049 Beijing, People's Republic of China}

\author[0000-0003-1675-1995]{Xiao Yang}
\affiliation{Key Laboratory of Solar Activity, National Astronomical Observatories, Chinese Academy of Sciences, 100101 Beijing, People's Republic of China}

\begin{abstract}

Duo to the large magnetic Reynolds number, the magnetic helicity originating from the solar interior can be carried away through the photosphere into the corona. However, the relationship between the accumulated magnetic helicity flux through the photosphere and the magnetic helicity in the corona is still unclear. By selecting 36 newly emerging active regions in the 23rd solar cycle, we apply optical flow methods to derive the accumulated magnetic helicity through the photosphere ($H_m^p$) by using the sequential longitudinal magnetograms, use nonlinear force-free field extrapolation to obtain the 3D coronal magnetic field, and adopt finite volume methods to calculate the instantaneous relative magnetic helicity in the corona ($H_m^c$) by using vector magnetograms. It is found that the local correlation tracking (LCT)-based $H_m^p$ is larger than $H_m^c$ in $1''$, and that the Differential Affine Velocity Estimator-based $H_m^p$ is more consistent with $H_m^c$ than the LCT-based $H_m^p$. $H_m^p$ is more consistent with $H_m^c$ in evaluation from $2''$ than from $1''$. Moreover, $H_m^c - H_m^p$ systematically shows consistency with the Hemispheric Helicity Rule (over 55\%), no matter which resolution and method are used. These estimations suggest that the consistency of $H_m^c$ and $H_m^p$ is partly dependent on the resolution of the magnetograms and the calculation methods.

\end{abstract}

\keywords{Solar Corona (1483); Solar Photosphere (1518); Solar Active Region Magnetic Fields (1975); Solar Magnetic Flux Emergence (2000)}


\section{Introduction}           
\label{sect:intro}

In 1833, Gauss discovered a simple formula for counting the algebraic intersection number when he was studying asteroid orbits. The Gauss linking number is actually a double integral, and it have been proved to be fundamental in knot theory, general topology, and modern topological field theory \citep{Ricca2011}. Supposing that the magnetic field can be approximated as an infinite number of individual field lines with infinitesimal flux, the magnetic helicity ($H_m= \int_{V} \bm{A} \cdot \bm{B} dV $; where $\bm{A}$ is the vector potential for the magnetic field $\bm{B}$) can be regarded as the sum of the connection numbers between any two magnetic field lines in a volume \citep{Moffatt1969}.

In a simply connected volume bounded by a magnetic surface ($\bm{B}_n|_S=0$), the magnetic helicity is gauge invariant in the transformations $\bm{A}$ to $\bm{A} + \nabla \psi$ \citep{Elsasser1956,Woltjer1958}. This invariance is untenable in a volume that is not bounded by a magnetic surface, such as the solar corona. However, a relative measure of the helicity of a simply connected volume with open field lines does exist \citep{Taylor1974,Berger1984,Berger1999}.

The sources of helicity are related to the solar dynamo, and the helicity also affects the solar cycle and solar activity in several ways. In the solar interior, the regeneration of the magnetic field and the 11-year solar cycle are both related to helicity \citep{Low2001,Vishniac2001,Choudhuri2003,Zhang2010}, while excessive magnetic helicity would also suppress the alpha effect of the solar dynamo. In the solar corona, the generation of magnetic helicity is more likely to lead to intense solar activities \citep{Low1996,Romano2011}. However, these two parts of the magnetic helicity are not equivalent, as the helicity generated by the solar dynamo may not lead to the same helicity in the corona, both in terms of the sign or the numerical value. \citet{Seehafer1990} first statistically analyzed the sign of the magnetic helicity in solar active regions, estimating the current helicity of 16 active regions and discovering that the helicity is predominantly negative in the northern hemisphere and positive in the southern hemisphere, which is called the ``Hemispheric Helicity Rule'' (HHR). Some of the following observations and statistical studies established that HHR is a relatively weak statistical trend that is satisfied by 60\%--75\% of active regions \citep{Pevtsov1995,Abramenko1996,Bao1996}, by 80\%--87\% of sigmoids \citep{Rust1996,Lim2009}, by 76\%--82\% of filaments \citep{Rust1994,Pevtsov2003}, by magnetic clouds \citep{Leamon2002}, and so on.

At present, there are three main kinds of methods for estimating magnetic helicity quantitatively: the optical flow methods, the finite volume methods, and the discrete flux tube methods \citep{Georgoulis2012,Valori2016,Guo2017}. The former two methods can be applied to calculate the accumulated magnetic helicity flux through the photosphere ($H_m^p$) and the instantaneous relative magnetic helicity in the corona ($H_m^c$), respectively.

The optical flow methods estimate the velocity field and magnetic vector potential to derive the integration of the magnetic helicity flux through the photosphere. The photospheric magnetic helicity flux is a function of time, and adopts a time integration to approximate the total coronal magnetic helicity (the latter being estimated by the finite volume methods). \citet{Chae2001} first calculated the magnetic helicity of an active region from a set of magnetograms by applying local correlation tracking (LCT) and Fourier transform (FT) methods. By introducing the flux transport velocity, \citet{Demoulin2003} suggested that all of the fluxes are tangential fluxes determined by the tracking methods. \citet{Schuck2006,Schuck2008} and \citet{Liu2012} developed some new optical flow methods (the Differential Affine Velocity Estimator (DAVE) and the Differential Affine Velocity Estimator for Vector Magnetograms) for calculating the magnetic helicity flux, by tracking magnetic footpoints. However, the helicity from the optical flow methods is a time integration, which lacks a natural ``reference'' level (see the details in Section~\ref{sect:SampSelect}).

Calculating the magnetic helicity in the corona by finite volume methods is more complicated than estimating the magnetic helicity flux through the photosphere by optical flow methods, and the significant point of the finite volume methods is how to obtain coronal magnetic fields. \citet{Cargill2009} pointed out the difficulties in the traditional approach of using the Zeeman effect to measure coronal magnetic fields. Some researchers have attempted to explain the observations of coronal oscillations by using magnetohydrodynamic (MHD) wave models, and the magnetic field strength of coronal magnetic fields can be inferred in this process \citep{Nakariakov1999,Tomczyk2007}. Another approach involves using sophisticated numerical computations to reconstruct realistic 3D nonlinear force-free coronal magnetic field models, which need observed photospheric vector magnetograms as boundary conditions \citep{Wiegelmann2012,Wiegelmann2021}. These coronal models include the MHD models, the magnetohydrostatics (MHS) models, the nonlinear force-free field (NLFFF) models, and the potential-field source-surface models. The force-free approach is a static model that neglects time-dependent phenomena and plasma flows, and the gradient of the plasma pressure and the gravity force in the MHS equation are also neglected as these terms are at least four orders of magnitude smaller than the magnetic pressure in the corona, so that the Lorentz force is vanishing and the electric currents are parallel/antiparallel to the magnetic field lines \citep{Wiegelmann2017}. However, whichever numerical computation method is chosen to approximate the 3D coronal magnetic field, they all have limitations. \citet{DeRosa2015} compared the influence of spatial resolution on the NLFFF model, and they found that the output of the NLFFF model depends on the spatial resolution of the input photospheric magnetograms (bottom boundary), with the values of free energy generally tending to be higher with increasing resolution, while the values of the relative magnetic helicity vary significantly with different resolutions. They also found that the results from the more highly resolved data were more self-consistent.

Although the NLFFF model does not represent the ground truth of coronal magnetic fields perfectly, this model still has a wide application range. \citet{Gary2001} constructed a simple 1D model for the magnetic stratification of the solar atmosphere, and constrained the magnetic and plasma pressures of this model using numerous observations at various heights. At a range of heights between the chromosphere and about 100 Mm, where the plasma $\beta \gg 1$ (Figure 3 in \citeauthor{Gary2001} \citeyear{Gary2001}), the NLFFF model is applicable. As such, we should focus on the variance of the observed photosphere magnetograms in preprocessing, and how well the force-free and divergence-free conditions are fulfilled after preprocessing (see the details in Section~\ref{sect:Merit}).

The solar corona is not a magnetically bounded volume; there is magnetic field emergence from the solar interior through the photosphere, the shedding of the magnetic field via the solar wind, and probably magnetic flux transportation between neighboring active regions. \citet{Berger1984} and \citet{Finn1985} proposed the concept of relative magnetic helicity for calculating the magnetic helicity in the corona by comparing a given field with the corresponding potential field (the potential field means zero current and the lowest-energy state). The relative magnetic helicity $H_R$ is given by the Finn--Antonsen formula, which is gauge invariant in open and multiply connected volumes (see the details in Section~\ref{sect:FiniteVol}).

The finite volume method can provide the relative magnetic helicity value of a 3D vector magnetic field in a bounded volume. The output of the finite volume method (relative magnetic helicity, magnetic energy) is usually at a certain moment. \citet{Yang2013,Yang2018} developed a finite volume method \footnote{https://sun10.bao.ac.cn/NAOCHSOS/rmhcs.htm} based on a Coulomb gauge, which is adopted in this paper. \citet{Valori2016} provided a systematic comparison of six finite volume methods\citep{Thalmann2011,Valori2012,Yang2013,Amari2013,Rudenko2014,Moraitis2014} for estimating the magnetic helicity, and they found that the spread in the computed values obtained by five of the six methods was only 3\%, which indicates the reliability and consistency of the finite volume methods.

Even though there are many studies of magnetic helicity calculations \citep{Zhang2008,Park2010,Thalmann2021}, the relationship between $H_m^p$ and $H_m^c$ is not yet clear in observations; for instance, the proportion of the magnetic helicity transported from the photosphere into the corona. In this paper, we utilize optical flow methods to derive the accumulated magnetic helicity through the photosphere ($H_m^p$), then apply the NLFFF model and the finite volume method to calculate the relative magnetic helicity in the corona ($H_m^c$), in order to estimate the amount of magnetic helicity transported through the photosphere into the corona.

The rest of this article is organized as follows. Section~\ref{sect:Data} shows the data sets and presents the criteria for selecting the samples and the preprocessing of the magnetograms. Section~\ref{sect:Met} introduces two optical flow methods, nonlinear force-free extrapolation, and the finite volume method for calculating $H_m^p$ and $H_m^c$, and shows the reliability of the NLFFF model. Section~\ref{sect:Res} shows the results of the comparison and the merit of the NLFFF model, and analyzes some reasons for the discrepancy between $H_m^p$ and $H_m^c$. A summary is then given in Section~\ref{sect:Sum}.

\section{Data and Preprocessing}
\label{sect:Data}

\subsection{MDI Data}

Considering the consistency and stability of the magnetograms, the full-disk line-of-sight magnetograms of the photosphere from the Michelson Doppler Imager (MDI; \citeauthor{Scherrer1995} \citeyear{Scherrer1995}) on board the Solar and Heliospheric Observatory are available for calculating the velocity field and $H_m^p$. There are two kinds of full-disk MDI magnetogram data: 1 minute cadence data and 96 minute cadence data. We select the 96 minute cadence data for the higher signal-to-noise ratio in this paper. MDI full-disk magnetograms are recorded by a 1024 $\times$ 1024 pixel CCD detector with a pixel size of $2''$, and the optical resolution of the magnetogram is $1.17''$ .

The effect of differential rotation needs to be removed with the following expression:
\begin{equation}
    \omega(\phi) = a + b\sin^2\phi + c\sin^4\phi,
\label{equ1}
\end{equation}
where $\phi$ presents the latitude, taking rotation coefficients with $a = 14.326$ deg day$^{-1}$, $b = -2.119$ deg day$^{-1}$, and $c = -1.832$ deg day$^{-1}$ \citep{Howard1990}. In addition, there is a coefficient $1/\cos\psi$ between the longitudinal field and the line-of-sight field for the geometrical foreshortening, where $\psi$ is the heliocentric angle of the region \citep{Liu2006}.

\subsection{SMFT Data}

The Solar Magnetic Field Telescope (SMFT) is a narrow-band filter-type magnetograph installed at the Huairou Solar Observation Station (HSOS), National Astronomical Observatories, Chinese Academy of Sciences. This telescope has been running for more than 35 years, and has obtained a large number of photospheric vector magnetograms. This filter measures at two wavelengths (the line center and $-75$ m\AA\ offset), with a bandwidth of 125 m\AA\ across the photospheric Fe\,{\small I} $\lambda$5324.19 spectral line. The spatial resolution is $0.38''$ and the field of view is around $6'\times 4'$ (there has been a little change over time). The photospheric vector magnetograms can be obtained from Stokes $\emph{I}$ , $\emph{Q}$ , $\emph{U}$, and $\emph{V}$ signals, using linear calibration with a proper calibration coefficient \citep{Ai1982,Ai1986,Su2004,Su2007}.

\subsection{Sample Selection and Data Preprocessing}
\label{sect:SampSelect}

\begin{figure}[!htbp]
   \centering
   \includegraphics[width=0.5\textwidth, angle=0]{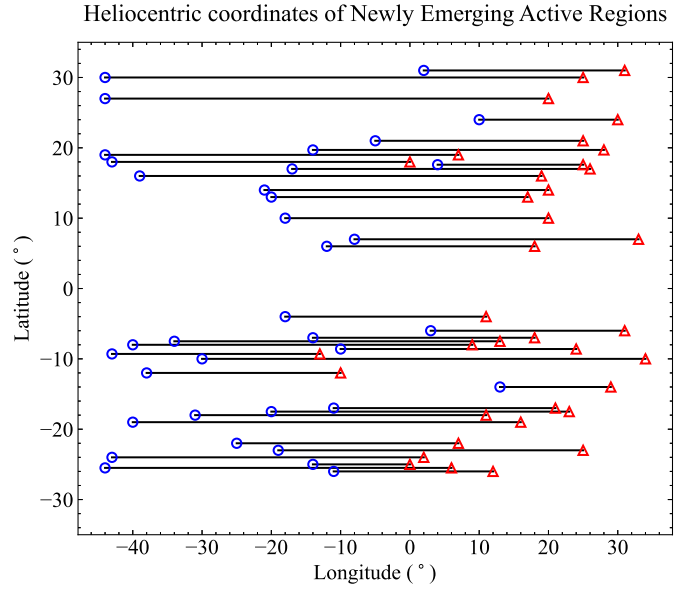}
   \caption{The heliocentric coordinates of 36 NEARs at the time of calculating $H_m^c$. The blue circles and red triangles represent the positions of the NEARs on the solar disk at the beginning and end times of the magnetic helicity transportation calculation, respectively.}
   \label{Fig1}
\end{figure}

We select 36 newly emerging active regions (NEARs) in the 23rd solar cycle as the sample, with 20 NEARs in the southern hemisphere and 16 in the northern hemisphere. Figure~\ref{Fig1} shows the heliocentric coordinates of the 36 NEARs during the calculation of $H_m^p$. The $H_m^p$ and $H_m^c$ of the NEARs are considered to be zero at the initial time ($t = 0$), thus the $H_m^p$ ($H_m^p = \int^{T}_{0} \frac{dH}{dt} dt$) and $H_m^c$ ($H_{m,t=T}$) can be compared at the given time $T$ (the moment calculating the relative magnetic helicity). The initial emerging time for a NEAR is defined as the moment when the magnetic flux started to increase from its quiet state, and in practice we adopt the time of the first MDI magnetogram for the same day of the flux emerging. The longitudes of all of the NEARs at the initial time and the given time $T$ are less than 35$^{\circ}$, to reduce the influence of projection effects, and the maximum longitudinal magnetic fluxes of all of the NEARs are greater than $10^{21}$ Mx.

\begin{figure}[!htbp]
   \centering
   \includegraphics[width=0.5\textwidth, angle=0]{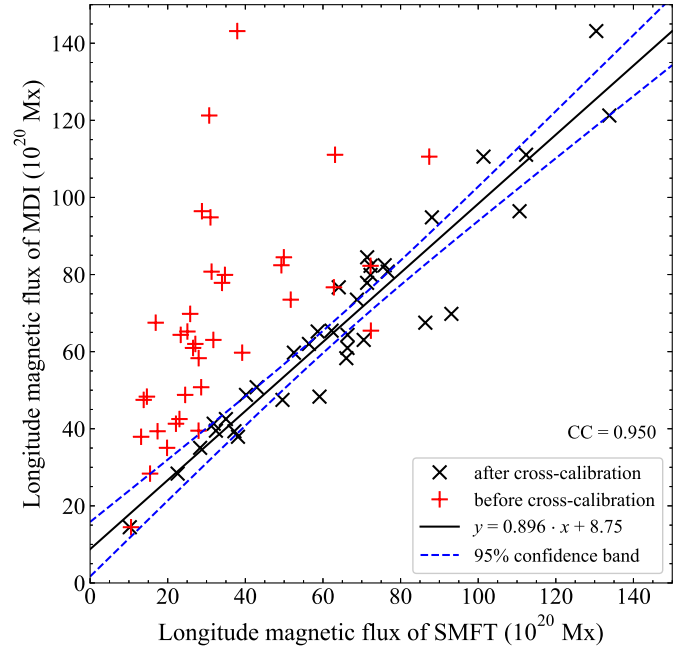}
   \caption{The longitudinal magnetic flux of SMFT and MDI magnetograms of 36 NEARs. The red and black dots represent the relations between two longitudinal magnetograms before and after cross-calibration. The black solid line represents the regression line, and the area between the two blue dashed lines is the 95\% confidence band.}
   \label{Fig2}
\end{figure}

\begin{figure}[!htbp]
    \centering
    \includegraphics[width=1.0\textwidth, angle=0]{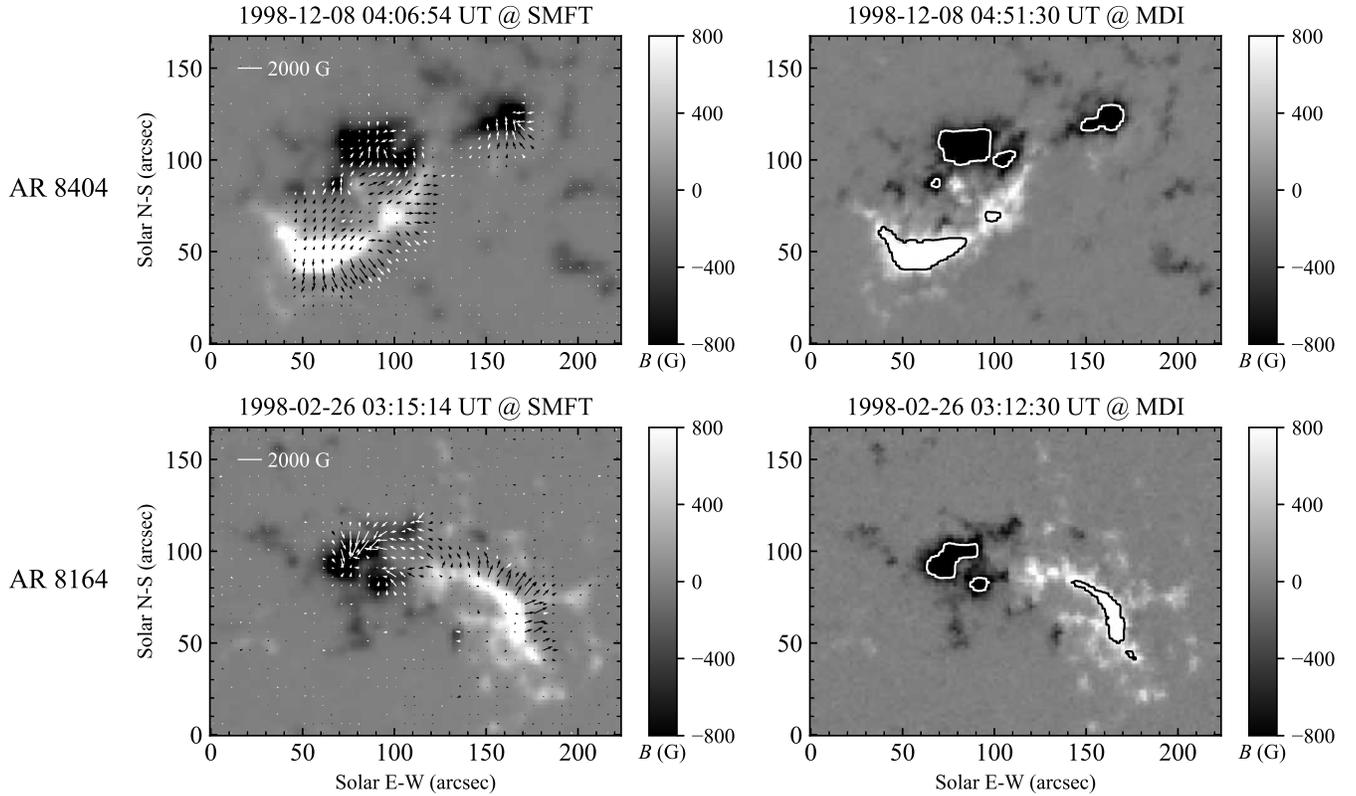}
    \caption{The magnetograms of SMFT and MDI (AR 8164 and AR 8404). The left panels represent the vector magnetograms from SMFT, and the right panels show the longitudinal magnetograms from MDI, contoured by $\pm800$ Gauss longitudinal fields of SMFT.}
    \label{Fig3}
    \end{figure}

There are systematic differences between the MDI and SMFT magnetograms, so a pixel-to-pixel cross-calibration is necessary for each pair of longitudinal magnetograms. Firstly, we resize the SMFT and MDI magnetograms to $1''$ per pixel, and the $2''$ reduced data are resized in the same way. Then we rotate the SMFT longitudinal magnetograms to align with the MDI magnetograms, and the SMFT transversal magnetograms are rotated following the longitudinal magnetograms. Finally, the slope of the linear fitting line between the two longitudinal magnetic fluxes is obtained from each pair of longitudinal magnetograms, and this slope is multiplied to the vector fields of SMFT to ensure consistency. The slopes of all 36 NEARs are presented in column 6 of Table~\ref{Tab2}. Due to the nonlinear relationship between the two instruments, the slopes range from 0.86 to 5.11. Figure~\ref{Fig2} shows the total longitudinal magnetic flux of the SMFT and MDI magnetograms of 36 NEARs, where the correlation coefficient is 0.950 after cross-calibration. Figure~\ref{Fig3} shows the magnetograms of NEAR 8164 and NEAR 8404 after preprocessing, indicating that the configurations of the longitudinal magnetic fields from the two instruments are similar. Furthermore, to evaluate the influence of different longitudinal fields on the subsequent helicity calculation and the quality of the data preprocessing, we introduce combined vector magnetograms, which are a combination of the SMFT transversal field and the MDI longitudinal field. After the above preprocessing, we can further calculate $H_m^p$ and $H_m^c$ at the given time $T$.

More specifically, some studies have been carried out to compare the performance of SMFT and other similar instruments \citep{Xu2007,Xu2016,Wang2009,Bai2014}. These studies indicate that the current calibration coefficients of SMFT are underestimated, especially for the transverse component. Since the transverse field is multiplied by the same slope as the longitudinal field, the underestimated calibration coefficients may become sources of discrepancies in the subsequent NLFFF model and the calculation of helicities.

\section{Methods}
\label{sect:Met}

\subsection{Optical Flow Methods for Calculating Magnetic Helicity Flux}

The magnetic helicity flux across an open boundary in the ideal HMDs is given by \citet{Berger1984} as
\begin{equation}
\left.\frac{dH}{dt}\right|_S= 2 \int_{S}(\bm{A}_p \cdot \bm{B}_t)V_n dS - 2 \int_{S}(\bm{A}_p \cdot \bm{V}_t)B_n dS,
\label{equ2}
\end{equation}
where $\bm{B}$ is the magnetic field, $\bm{A}_p$ is the vector potential of the current-free field $\bm{B}_p$, $\bm{V}$ is the plasma velocity on the photosphere, and the subscripts ``$t$'' and ``$n$'' denote the tangential and normal components, respectively. The first term of the right-hand side, called the emergence term, represents the magnetic helicity being transported through a surface, while the second term, called the shear term, represents the horizontal shuffling motion of field lines on this surface. \citet{Demoulin2003} introduced the flux transport velocity $\bm{U}$ ($\bm{U}=\bm{V}_t - \frac{V_n}{B_n}\bm{B}_t$), so Equation (\ref{equ2}) becomes
\begin{equation}
    \left.\frac{dH}{dt}\right|_S= - 2 \int_{S}(\bm{A}_p \cdot \bm{U}_t)B_n dS.
\label{equ3}
\end{equation}
And they further pointed out that, for both energy and helicity, all of the fluxes are present in the tangential fluxes determined by the tracking methods. Therefore, the magnetic helicity flux could be derived from longitudinal magnetograms, since the injection by vertical motions can be neglected.

From Equation (\ref{equ3}), the flux transport velocity and vector potential are necessary for calculating the magnetic helicity flux. \citet{Chae2008a} evaluated three optical flow techniques (LCT: \citeauthor{November1988} \citeyear{November1988}; DAVE: \citeauthor{Schuck2006} \citeyear{Schuck2006}; and the Nonlinear Affine Velocity Estimator: \citeauthor{Schuck2005} \citeyear{Schuck2005}) to obtain the transverse velocity field. We apply the first two methods to derive $H_m^p$ in this paper. $\bm{A}_p$ is the vector potential of the current-free field $\bm{B}_p$ with the gauge $\nabla \cdot \bm{A}_p = 0$, which can be computed based on the observed magnetograms ($B_z$), and the FT is an efficient approach for deriving $\bm{A}_p$ \citep{Chae2001}.

LCT is a widely used method for detecting the motion of small fractions between two successive magnetograms \citep{Leese1970,November1988}. The physically significant displacement of the same structure between two successive magnetograms should be larger than the pixel size and smaller than the apodizing window size at the same time. The significant transverse velocity of photospheric fluxes is less than 1.5 km s$^{-1}$, the possible shift between two successive MDI magnetograms is less than 11.91 pixels (8703.6 km), so the FWHM is chosen as 12 pixels, considering the characteristic time scale of structural evolution of the magnetic fields \citep{Chae2001}.

The relationship between the FWHM and Gaussian window scale $\sigma$ is $\sigma \approx \textrm{FWHM}/2.35$, and the relationship between the Gaussian window scale $\sigma$ used by LCT and the absolute window width $w_0$ implemented in the differential techniques is $\sigma \approx 0.39 w_0$ \citep{Schuck2006}. For consistency with LCT, the window size is set as 13 pixels in DAVE.

\subsection{NLFFF Extrapolation for Reconstructing the Coronal Magnetic Field}

THe equilibrium structure of a magnetic field and plasma can be depicted by the MHS equation
\begin{equation}
  - \nabla p + \frac{1}{{\mu}_0}(\nabla \times \bm{B}) \times \bm{B} - \rho \bm{g} = 0,
\label{equ4}
\end{equation}
where $p$ is the pressure, $\rho$ is the density, $\bm{g}$ is the gravitational acceleration, and ${\mu}_0$ is the vacuum permeability. The magnetic pressure is about four orders of magnitude higher than the plasma pressure in the corona \citep{Wiegelmann2017}, and all nonmagnetic forces---like pressure gradients and gravity---can be neglected, so Equation (\ref{equ4}) can be simplified to
\begin{equation}
 (\nabla \times \bm{B}) \times \bm{B} = 0,
\label{equ5}
\end{equation}
in which the magnetic field is parallel to the current, and the Lorentz force is zero, so it is called a ``force-free equation.'' The force-free fields are characterized by Equation (\ref{equ5}) and the divergence-free equation ($\nabla \cdot \bm{B} = 0$). We adopt a nonlinear extrapolation method \citep{Wiegelmann2004} based on the assumption of a force-free field to deduce the configuration of the coronal magnetic field.

The measurements of the photospheric magnetic field contain inconsistencies and noise, especially for the transversal components. Furthermore, the photospheric magnetic field is not strictly force-free and often not consistent with the assumption of a force-free field above the magnetogram. So the observed data need to be revised toward suitable boundary conditions for nonlinear force-free extrapolation. Before extrapolation, all of the vector magnetograms from SMFT are preprocessed according to a preprocessing procedure \citep{Wiegelmann2006}.

\subsection{Finite Volume Method for Calculating Relative Magnetic Helicity}
\label{sect:FiniteVol}

The magnetic helicity is a strictly conserved quantity in ideal MHDs, and it is still approximately conserved even in nonideal conditions; for example, in the corona, with large magnetic Reynolds numbers ($10^{8}$--$10^{15}$) and moderate resistivity \citep{Taylor1974}.

The relative magnetic helicity \citep{Berger1984} in a finite volume is given by the Finn--Antonsen formula \citep{Finn1985}:
\begin{equation}
    \textsl{H}_{R}=  \int (\bm{A} + \bm{A}_p) \cdot (\bm{B} - \bm{B}_p) d^{3}x,
\label{equ6}
\end{equation}
where $\bm{B}_p$ and $\bm{A}_p$ represent the potential field and vector potential, respectively. This concept provides an alternative approach for measuring the magnetic helicity in the corona.

\citet{Yang2013,Yang2018} developed a method for calculating the vector potentials of magnetic fields given in finite volumes with a balanced magnetic flux on all of the side boundaries. This method employs the Coulomb gauge, and uses a fast Laplace/Poisson solver to obtain the vector potentials for a given magnetic field and for the corresponding potential (current-free) field. The instantaneous relative magnetic helicity in the corona $H_m^c$ is derived through this method at the given time.

\subsection{The Merit of NLFFF Model Solutions}
\label{sect:Merit}

In order to compare the NLFFF model solutions $\bm{B}$ and the observed photospheric vector fields $\bm{b}$, the force-freeness of the magnetic fields, the degree of convergence toward the NLFFF model solution, and the solenoidal property of the NLFFF model solution should be evaluated before calculating the relative magnetic helicity by finite volume methods. Some metrics can be used to test the performance of the NLFFF modeling.

(i). According to \citet{Low1985}, the magnetic field is force-free when
\begin{equation}
 |F_x| \ll |F_p|, \quad |F_y| \ll |F_p|, \quad |F_z| \ll |F_p|,
\label{equ7}
\end{equation}
where $F_x$, $F_y$, and $F_z$ are the components of the net Lorentz force, and $F_p$ is the characteristic magnitude of the total Lorentz force. In this assumption, the magnetic field vanishes as $z$ goes to infinity, and the contribution to the net Lorentz force is just limited to the plane $z = 0$, therefore
\begin{equation}
\left.\begin{lgathered}
    F_x = -\frac{1}{4\pi} \int B_xB_zdxdy  \\
    F_y = -\frac{1}{4\pi} \int B_yB_zdxdy  \\
    F_z = -\frac{1}{8\pi} \int (B_z^2-B_x^2-B_y^2)dxdy  \\
    F_p = \frac{1}{8\pi} \int (B_z^2+B_x^2+B_y^2)dxdy,
\end{lgathered} \right.
\label{equ8}
\end{equation}
where $B_x$, $B_y$, and $B_z$ are the two components of the horizontal magnetic field and the vertical magnetic field, respectively. When the conditions in Equation (\ref{equ7}) are satisfied, the corresponding magnetic field can be regarded as being force-free. \citet{Metcalf1995} proposed that when $|F_x/F_p|$, $|F_y/F_p|$, and $|F_z/F_p|$ are all less than 0.1, the magnetic field can be considered as a force-free field. The average values of these three metrics of observed photospheric vector fields are 0.101, 0.143, and 0.355; after the preprocessing of the NLFFF model, these metrics change to 0.005, 0.004, and 0.028. The numbers of NEARs that satisfy the criterion of \citet{Metcalf1995} are more than 30 for different resolutions.

(ii). The degree of convergence toward the NLFFF model solution can be evaluated by the integral measures of the Lorentz force and divergence \citep{Schrijver2006}:
\begin{equation}
\left.\begin{lgathered}
    L_f = \frac{1}{V} \int_{V} B^{-2} | (\nabla \times \bm{B}) \times \bm{B} |^2 dV \\
    L_d = \frac{1}{V} \int_{V} | \nabla \cdot \bm{B} |^2 dV \\
    L = L_f + L_d,
\end{lgathered} \right.
\label{equ9}
\end{equation}
where $L_f$ and $L_d$ measure how well the force-free and divergence-free conditions are fulfilled. In addition, the current-weighted average of the sine of the angle between the magnetic field and the electrical current density can also evaluate how well the force-free condition is satisfied:
\begin{equation}
    \sigma_J = \left( \sum_{i} \frac{|\bm{J}_i \times \bm{B}_i|}{B_i} \right) \left/ \sum_{i}|\bm{J}_i | \right..
\label{equ10}
\end{equation}
For a perfect force-free field, these four metrics result in $L_f = L_d = L = \sigma_{J} = 0$.

(iii). Two metrics can be used to evaluate the agreement of the solenoidal condition in the NLFFF model solution: the fractional flux $f_i$ and the magnetic energy of the nonsolenoidal component $E_{\textrm{div}}$. The expressions are given by \citep{Wheatland2000,Valori2013}
\begin{align}
    &f_i = \frac{\int_{\Delta S_i} \bm{B} \cdot \bm{dS}}{\int_{\Delta S_i} | \bm{B} | dS} \approx \frac{(\nabla \cdot \bm{B})_i \Delta V_i}{B_i A_i} \\
    &E_{\textrm{div}} = \int_{V} \frac{| \bm{B}_{\textrm{div}} \cdot \bm{B}_{\textrm{div}} |}{8 \pi} dV,
\label{equ11}
\end{align}
where $A_i$ is the surface area of the small volume, and $\bm{B}_{\textrm{div}}$ is the nonsolenoidal component of the NLFFF model solutions $\bm{B}$. For perfect divergence-free fields, $f_i = 0$ and $E_{\textrm{div}} = 0$. Moreover, it should be pointed out that one critical threshold ($E_{\textrm{div}}/E = 0.1$) is known so far for NLFFF models qualifying for subsequent helicity in volume computations \citep{Valori2013,Thalmann2019}. Out of the all 36 NEARs, 32 NEARs satisfy this threshold at $1''$, and 26 NEARs satisfy this threshold at $2''$.

In this paper, the grid number of extrapolation is $256 \times 200 \times 120$ for $1''$ reduced data and $128 \times 100 \times 60$ for $2''$ reduced data. The boundary layer is set to 16 and 8 for $1''$ and $2''$, respectively. The average values and standard deviations of these metrics for the 36 NEARs are listed in Table \ref{Tab1}, with the first and second lines showing the cases of $1''$ and $2''$, respectively. The proportions of the NLFFF models satisfying these quality levels are more than 70\% in these two resolutions. All of these metrics indicate that the force-freeness of the magnetic field is satisfied after the preprocessing, the convergence toward the NLFFF model solution is achieved, and the solenoidal condition is fulfilled. Therefore, the NLFFF model solution should be considered reliable in the finite volume magnetic helicity calculation. Moreover, the performance of NLFFF model at $1''$ is better than that at $2''$, from the metrics of Table \ref{Tab1}.

\begin{table}  \addtolength{\tabcolsep}{-3pt}
\begin{center}
\caption{Testing the Performance of the NLFFF Model Solution}
\label{Tab1}
\renewcommand\arraystretch{1.0}
 \begin{tabular}{cccccccc}
  \hline\hline\noalign{\smallskip}
 & $L_f$ & $L_d$ & $L$ & $\sigma_J$ & $f_i$ & $E_{\textrm{div}}/E$ \\
  \noalign{\smallskip}\hline\noalign{\smallskip}
$1''$ & $0.128\pm 0.011$ & $0.081\pm 0.007$ & $0.210\pm 0.018$ & $0.086\pm 0.003$ & $0.007\pm 0.000$ & $0.082\pm 0.006$ \\
$2''$ & $0.172\pm 0.018$ & $0.102\pm 0.010$ & $0.274\pm 0.028$ & $0.087\pm 0.003$ & $0.014\pm 0.000$ & $0.093\pm 0.007$ \\
  \noalign{\smallskip}\hline
\end{tabular}
\end{center}
\end{table}

\section{Results and Analysis}
\label{sect:Res}

Table \ref{Tab2} lists the calculation results of all 36 NEARs, including $F_\textrm{\tiny{SMFT}}$ and $F_\textrm{\tiny{MDI}}$ (the longitudinal magnetic fluxes at the given time), the slope between two longitudinal magnetograms from the SMFT and MDI data (see the detail in Section~\ref{sect:SampSelect}), $H_m^p$ (the integral magnetic helicity flux through the photosphere; columns 11--14), and $H_m^c$ (the instantaneous relative magnetic helicity in the corona; columns 7--10) and $H_m^c - H_m^p$ (column 15 lists examples of the difference between $H_m^c$ and LCT-based $H_m^p$ at $2''$). 16 NEARs are located in the northern hemisphere and 20 NEARs in the southern hemisphere. The subscripts ``1'' and ``2'' denote the resolutions of $1''$ and $2''$, respectively. The slopes (column 6) range from 0.86 to 5.11, and the average is 2.21, which originates from the nonlinear relationship between the two instruments.

\subsection{Effects of Different Resolutions}

First, the effects of the different resolutions ($1''$ and $2''$) on the individual methods (the NLFFF model, LCT, and DAVE) are mainly reflected in the signs and values of the helicity. For the $H_m^c$ calculated from the NLFFF model solution and the finite volume method, 35 of the 36 NEARs (97.22\%) have the same sign in different resolutions, and the absolute value of $H_m^c$ ($|H_m^c|$) decreases with the increase in the resolution in all NEARs. For the LCT-based $H_m^p$, 32 of the 36 NEARs (88.89\%) have the same sign in different resolutions, and the absolute value of $H_m^p$ ($|H_m^p|$) increases following the increase in the resolution in 28 NEARs (77.78\%). For the DAVE-based $H_m^p$, 35 of the 36 NEARs (97.22\%) have the same sign in different resolutions, and $|H_m^p|$ reduces with the increase in the resolution in 33 NEARs (91.67\%). These results indicate that the absolute values of the magnetic helicities vary following the change in the resolution, and that the estimation of LCT-based $H_m^p$ is more sensitive to resolution than the other two methods.

Second, the different resolutions affect the relation of $H_m^c$ and $H_m^p$, which is also reflected in signs and values. Comparing the $H_m^c$ and LCT-based $H_m^p$, 26 NEARs (72.22\%) have the same sign at $1''$ resolution, while 25 NEARs (69.44 \%) share the same sign at $2''$ resolution. In 28 NEARs (77.78\%), $|H_m^p|$ is closer to $|H_m^c|$ at $2''$ than $1''$. For the $H_m^c$ and DAVE-based $H_m^p$, 29 NEARs (80.56\%) have the same sign at $1''$ and 27 NEARs (75\%) share the same sign at $2''$. In 19 NEARs (52.78\%), $|H_m^p|$ is closer to $|H_m^c|$ at $2''$ than $1''$. These results indicate that the reduced spatial resolution is more likely to give space for the discrepancy of sign between $H_m^c$ and $H_m^p$, and is more likely to decrease the discrepancy between $|H_m^c|$ and $|H_m^p|$.

Third, the distribution of the net magnetic helicity ($H_m^c - H_m^p$) affects the comparison of $H_m^c$ and $H_m^p$.  $\Delta H_\textrm{\tiny{LCT}}$ is $H_m^c$ subtracted by LCT-based $H_m^p$, and $\Delta H_\textrm{\tiny{DAVE}}$ is $H_m^c$ subtracted by DAVE-based $H_m^p$. For the northern hemispherical NEARs, $\Delta H_\textrm{\tiny{LCT}}$ is negative in 13 NEARs (81\%) for $2''$ and 9 NEARs (56\%) for $1''$, and $\Delta H_\textrm{\tiny{DAVE}}$ is negative in 12 NEARs (75\%) for $2''$ and 10 NEARs (62\%) for $1''$. As for the southern hemispherical NEARs, $\Delta H_\textrm{\tiny{LCT}}$ is positive in 16 (80\%) NEARs for $2''$ and 11 NEARs (55\%) for $1''$, and $\Delta H_\textrm{\tiny{DAVE}}$ is positive in 11 NEARs (55\%) for both $2''$ and $1''$. The proportions complying with HHR are all over 55\%, and cases of $2''$ are better than $1''$. These results indicate that the net magnetic helicity in the corona at the initial time ($t = 0$) is in accordance with HHR, and that this part of the magnetic helicity in the corona will considerably affect the comparison between $H_m^c$ and $H_m^p$.

\subsection{Effects of Different Optical Flow Methods}

\begin{figure}[!htbp]
   \centering
   \begin{minipage}{\textwidth}
       \centering
       \includegraphics[width=0.4\textwidth, angle=0]{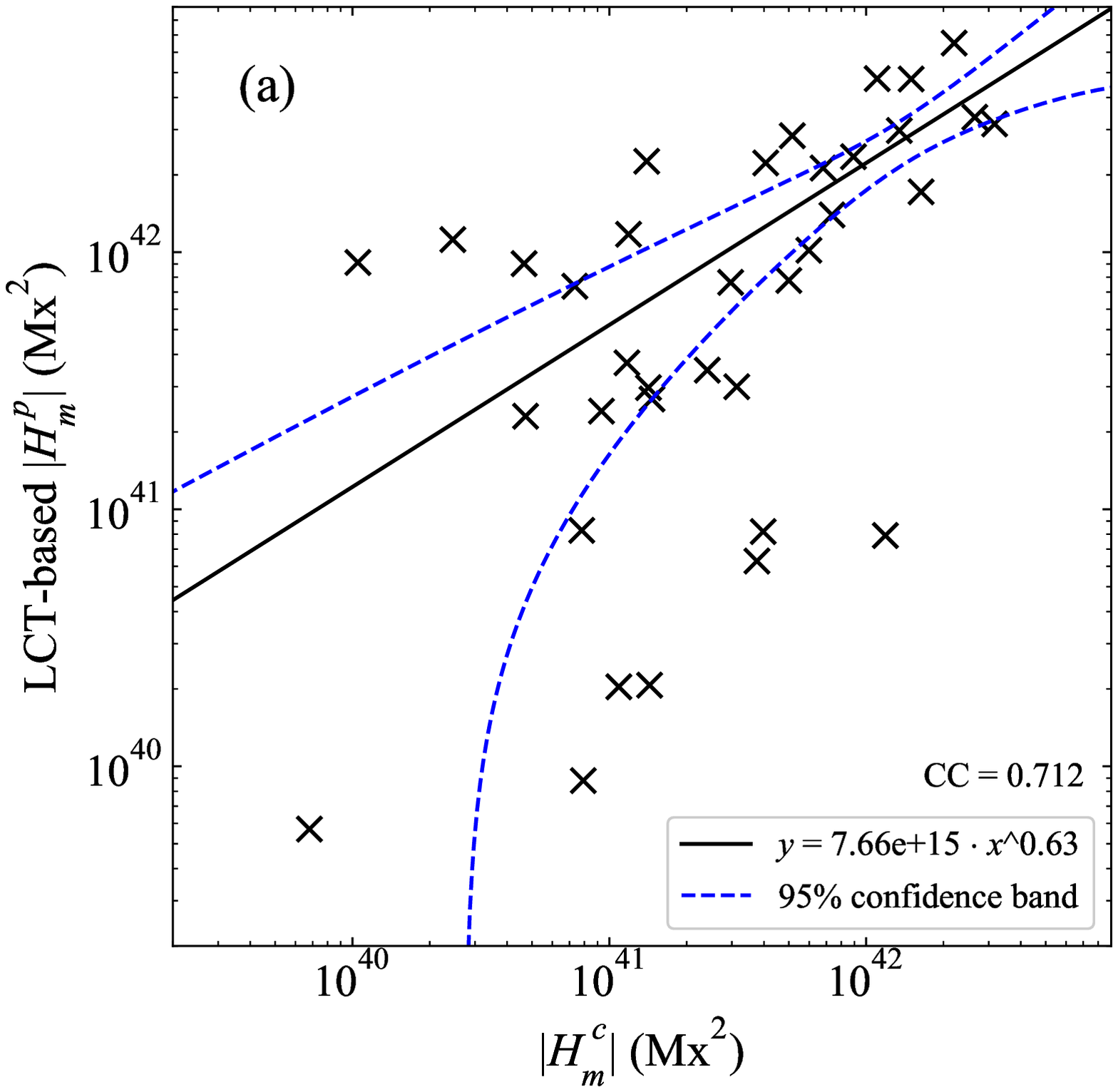}
       \hspace*{2mm}
       \includegraphics[width=0.4\textwidth, angle=0]{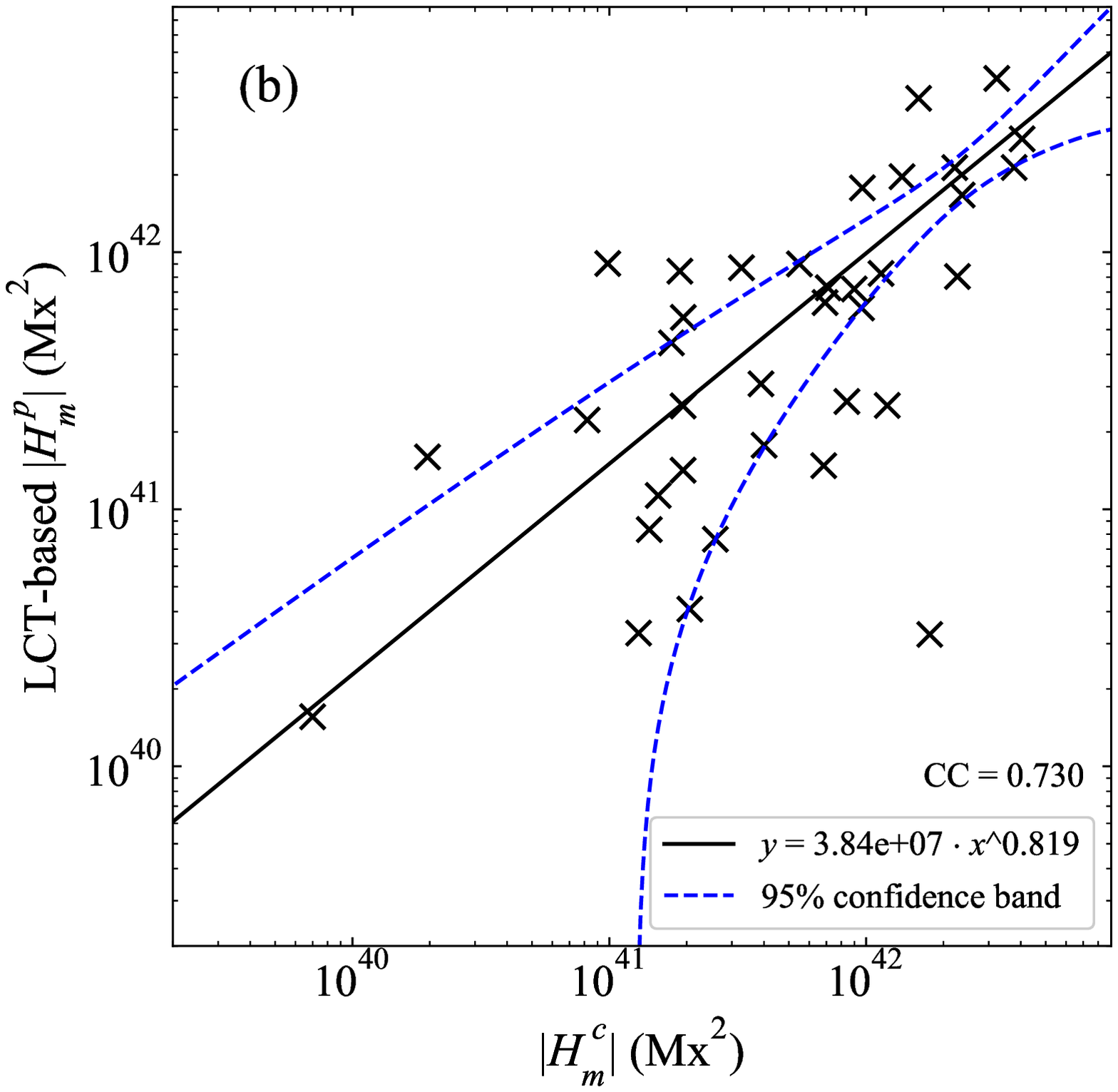}
       \vspace*{2mm}
   \end{minipage}
   \begin{minipage}{\textwidth}
       \centering
       \includegraphics[width=0.4\textwidth, angle=0]{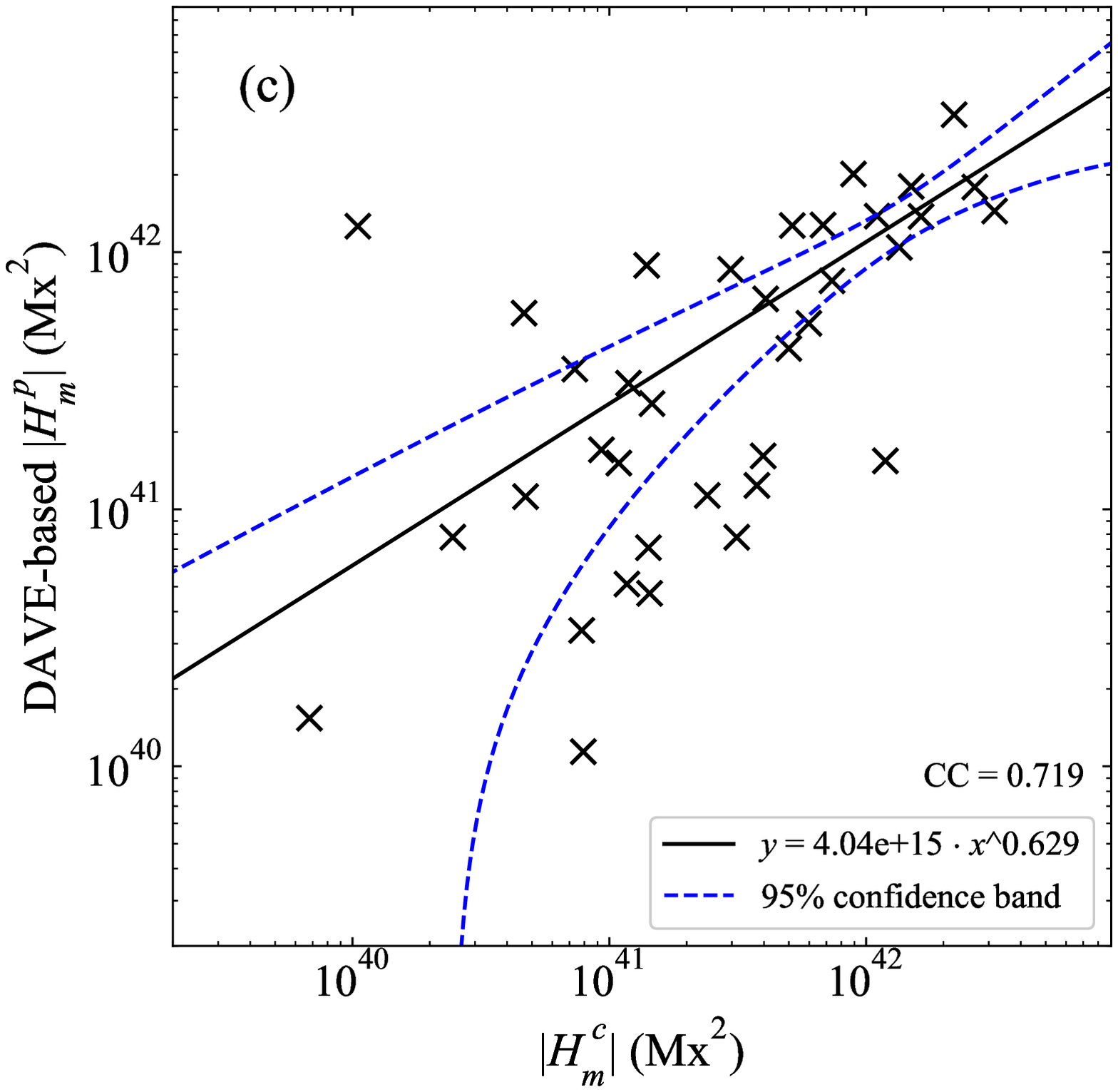}
       \hspace*{2mm}
       \includegraphics[width=0.4\textwidth, angle=0]{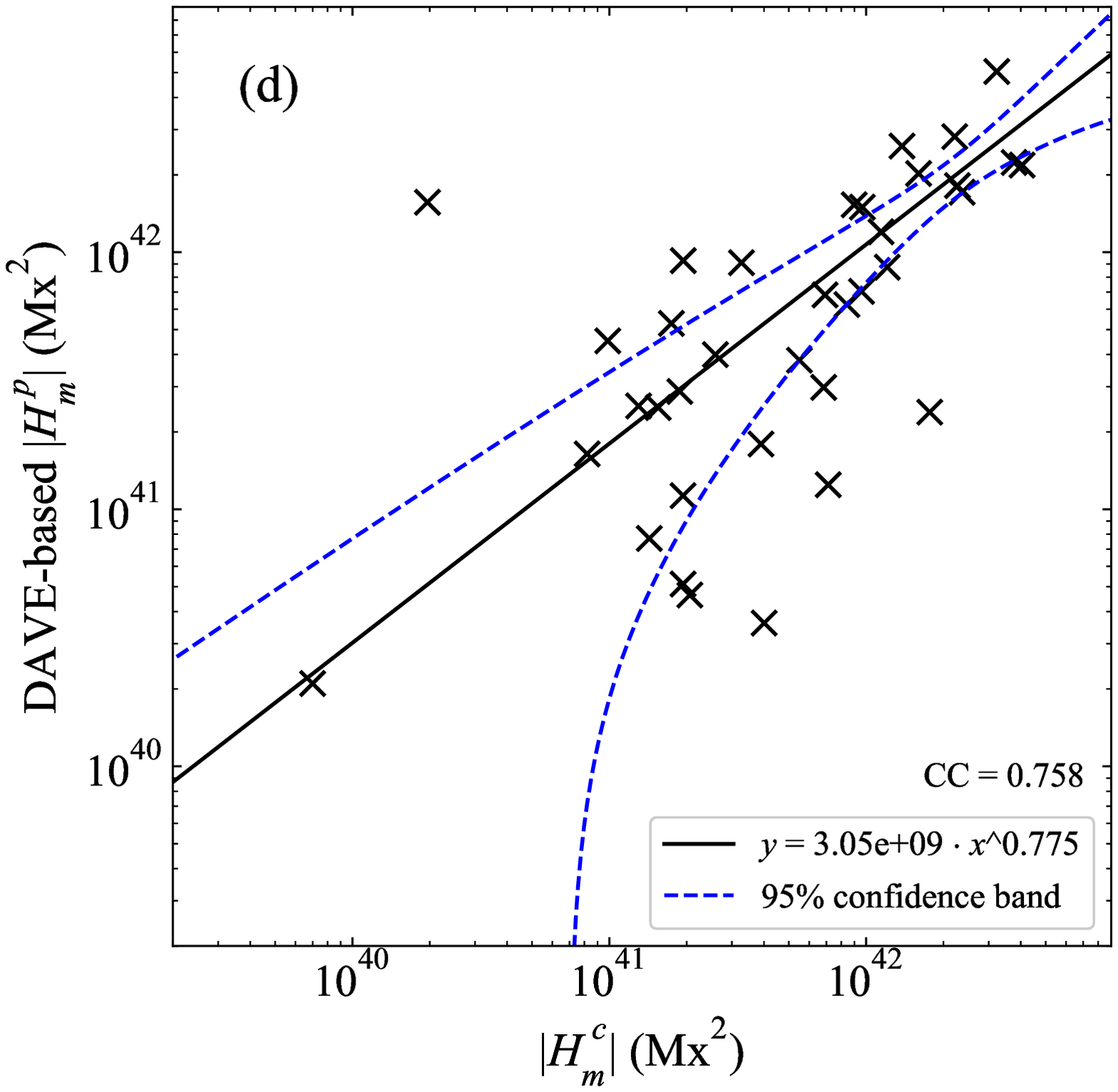}
       \vspace*{2mm}
   \end{minipage}
   \begin{minipage}{\textwidth}
       \centering
       \includegraphics[width=0.4\textwidth, angle=0]{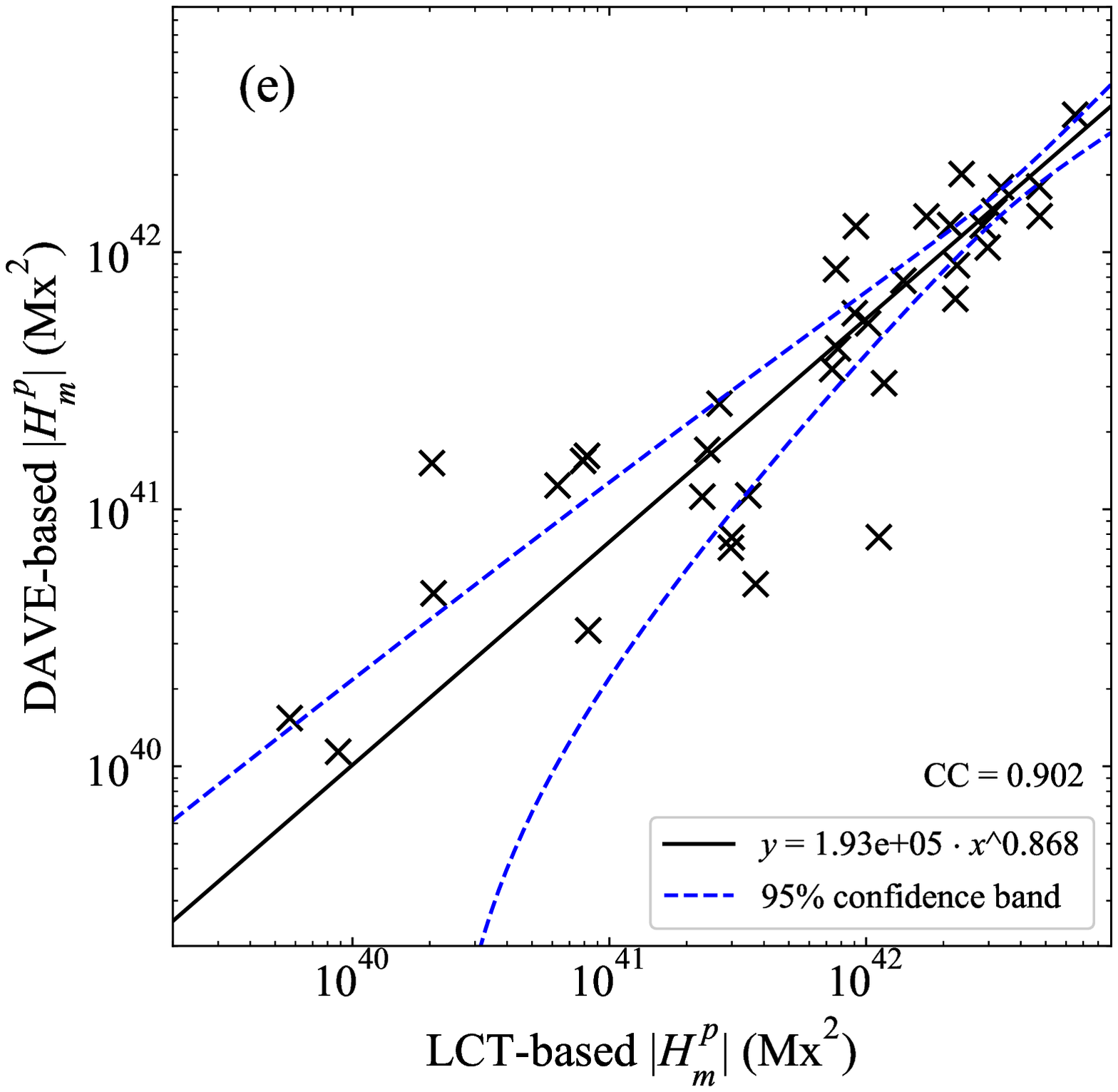}
       \hspace*{2mm}
       \includegraphics[width=0.4\textwidth, angle=0]{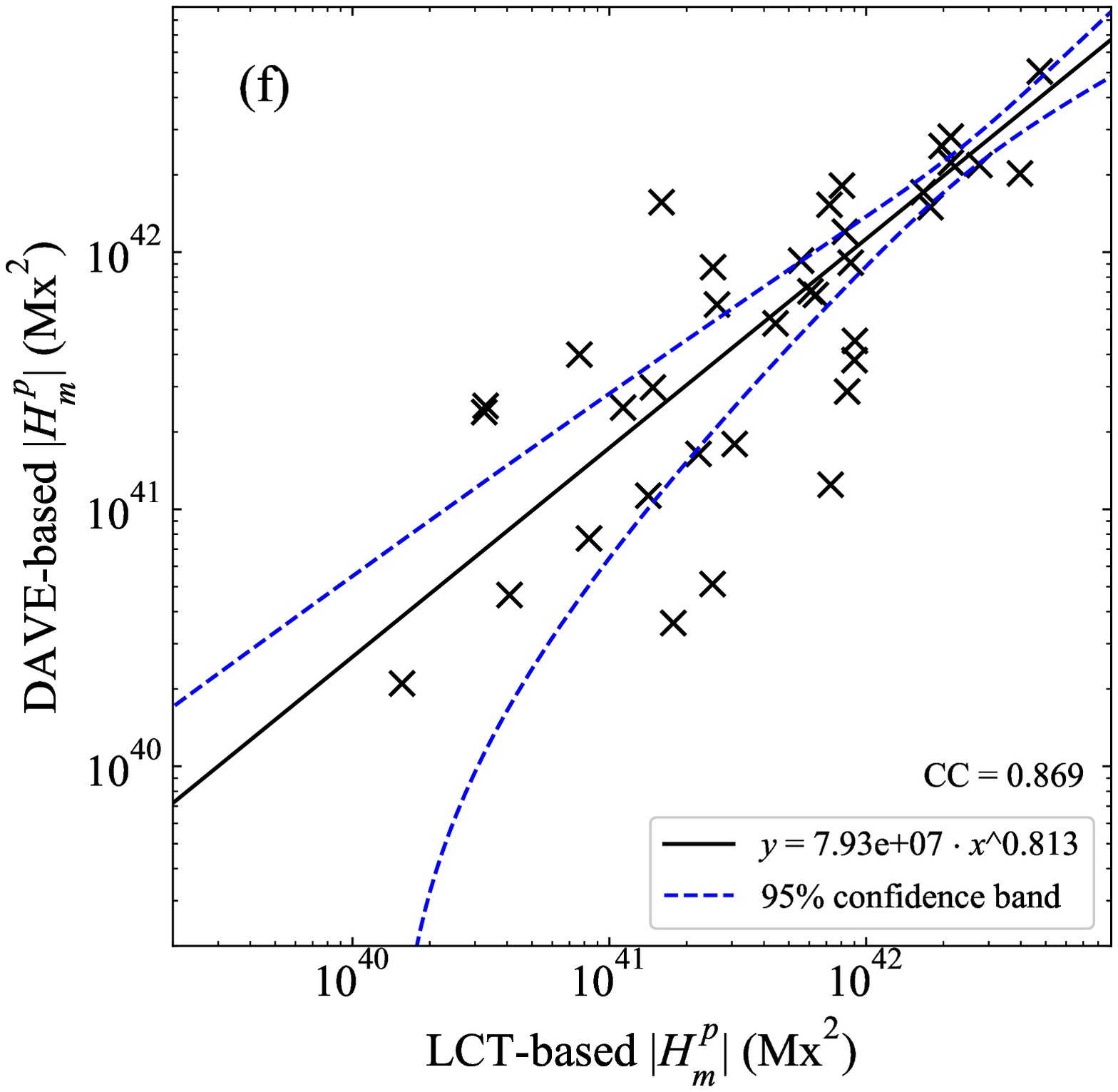}
   \end{minipage}

   \caption{The relation between $H_m^c$ and $H_m^p$ for the different optical flow methods. The left and right columns show the $1''$ and $2''$ reduced data, respectively.}
   \label{Fig4}
\end{figure}

Columns 11--14 of Table \ref{Tab2} list the $H_m^p$ from the different optical flow methods. Figure \ref{Fig4} shows the relation between $|H_m^c|$ and $|H_m^p|$ based on the different optical flow methods: the left and right panels represent $1''$ and $2''$, and the top, middle, and bottom panels show the relation between $|H_m^c|$ and LCT-based $|H_m^p|$, $|H_m^c|$ and DAVE-based $|H_m^p|$, and LCT-based $|H_m^p|$ and DAVE-based $|H_m^p|$, respectively. The coefficients and 95\% confidence intervals of the fitting lines are given for these figures.

At $1''$, the LCT-based $H_m^p$ have the same sign as the DAVE-based $H_m^p$ in 31 NEARs (86.11\%); the LCT-based $|H_m^p|$ is larger than the DAVE-based $|H_m^p|$ in 27 NEARs (75\%); and the DAVE-based $|H_m^p|$ is more consistent with the $|H_m^c|$ than the LCT-based $|H_m^p|$ in 27 NEARs (75\%). As for $2''$, the proportion with the same sign is 77.78\% (28 NEARs); the LCT-based $|H_m^p|$ is less than the DAVE-based $|H_m^p|$ in 23 NEARs (63.89\%); and the DAVE-based $|H_m^p|$ is closer to the $|H_m^c|$ in 19 NEARs (52.78\%). These results suggest that the DAVE-based $H_m^p$ is more consistent with $H_m^c$ than the LCT-based $H_m^p$ in whichever resolution.

From Figures \ref{Fig4}(a)--(d), the spread of the dots is smaller in the range of $|H_m^c| > 10^{42}$ Mx$^2$ than in the range of $|H_m^c| < 10^{42}$ Mx$^2$ for both of the two optical flow methods, which suggests that the consistency between $|H_m^p|$ and $|H_m^c|$ is better for the NEARs with larger magnetic helicity (longitudinal magnetic flux). The spread of dots from DAVE is smaller than from LCT for both of the two resolutions, which is consistent with the previous analysis.

\citet{Chae2007} studied the magnetic helicity injection for NOAA 10696 and pointed out that the helicity measurements using the DAVE method yield systematically higher values of helicity injection than those using the LCT method. In their estimation, the discrepancy is moderately small (less than 10\%). From Figure \ref{Fig4}(e) ($1''$), it is obvious that the DAVE-based $|H_m^p|$ is systematically larger than the LCT-based $|H_m^p|$ for the NEARS with less longitudinal magnetic flux (and thus less magnetic helicity), for which the corresponding plot symbols are above the hypothetical 1:1 line; while the former is systematically less than the latter for NEARs with larger longitudinal magnetic flux (and thus larger magnetic helicity), for which the corresponding plot symbols are below the hypothetical 1:1 line. However, there is no such trend for $2''$ (Figure \ref{Fig4}(f)). In our estimation, the average discrepancy between the LCT-based $H_m^p$ and the DAVE-based $H_m^p$ partly relates to the resolution, which is around 50\% for $1''$ and 7.3\% for $2''$.

\subsection{Effects of Different Longitudinal Fields}

\begin{figure}[!htbp]
   \centering
        \includegraphics[width=0.4\textwidth, angle=0]{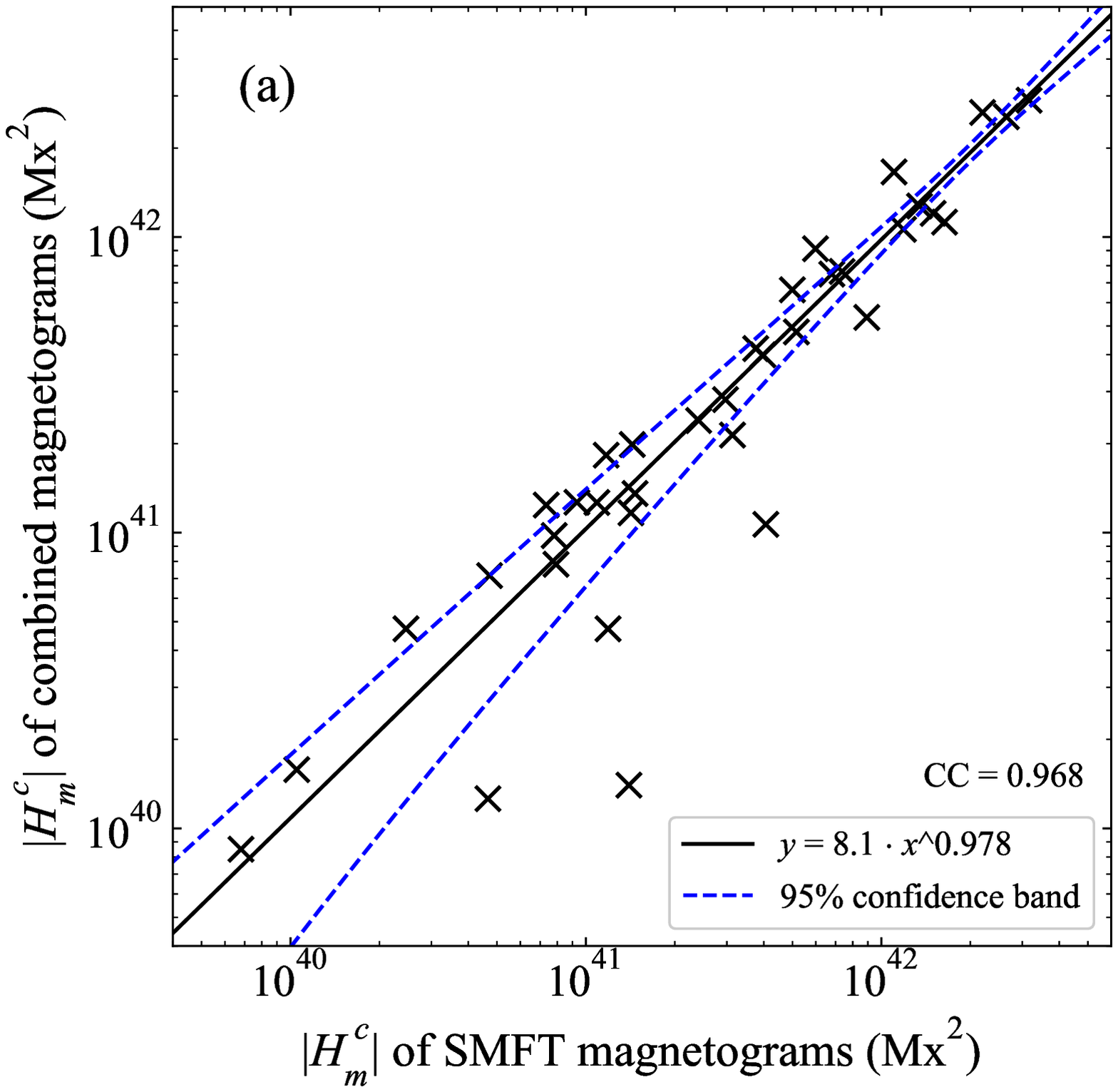}
        \hspace*{2mm}
        \includegraphics[width=0.4\textwidth, angle=0]{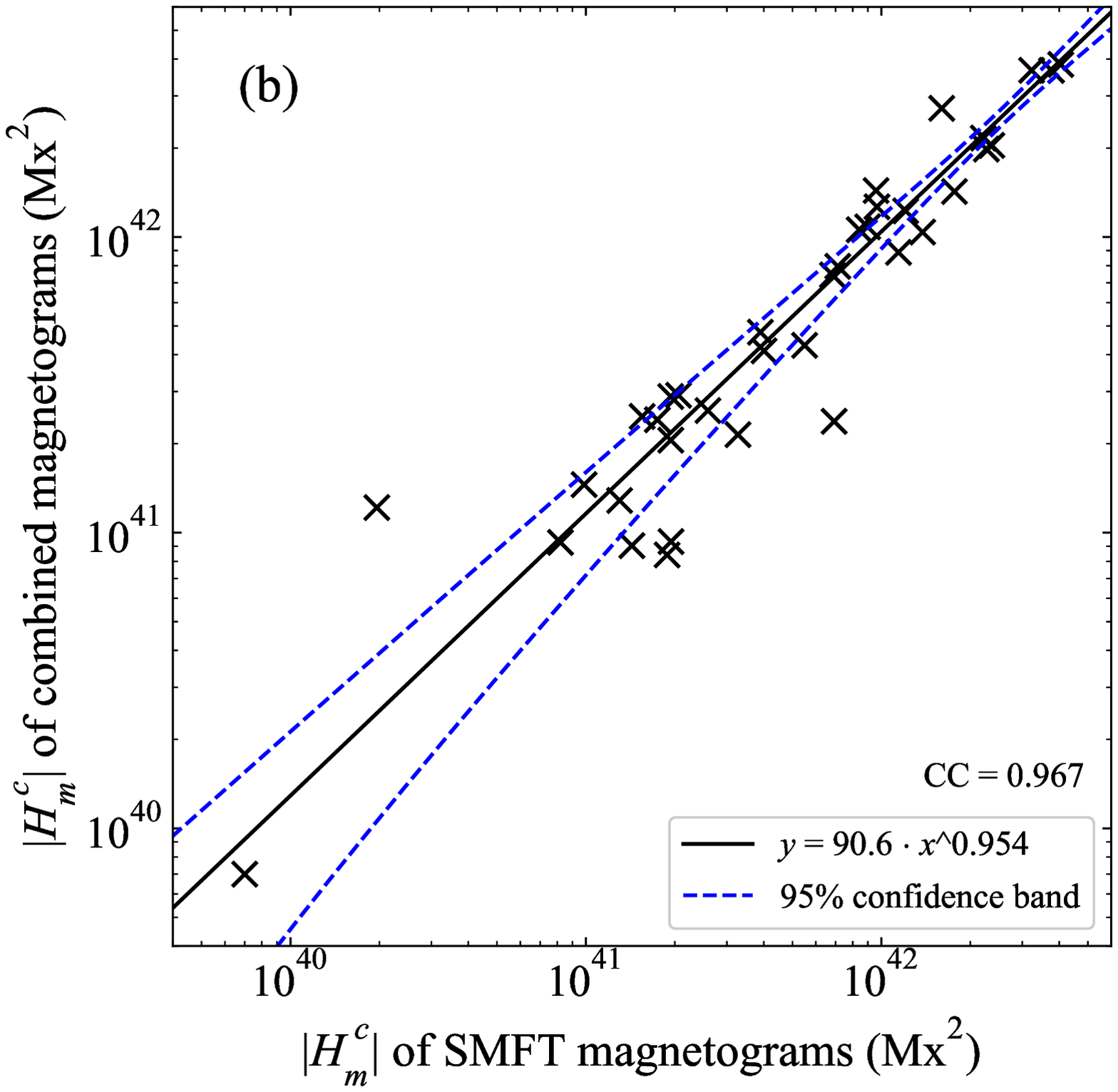}
   \caption{The relation of $H_m^c$ between SMFT and combined vector magnetograms. The left and right panels show the $1''$ and $2''$ reduced data, respectively.}
   \label{Fig5}
\end{figure}

Figure \ref{Fig5} shows the relation of $|H_m^c|$ between SMFT and combined vector magnetograms; the left and right panels represent $1''$ and $2''$. Columns 7--10 of Table \ref{Tab2} list the $H_m^c$ from the SMFT magnetograms and combined magnetograms. At $1''$, the signs of the different $H_m^c$ are the same in 32 NEARs (88.89\%), and the correlation coefficient of the different $|H_m^c|$ is 0.968. As for $2''$, the signs are the same in 35 NEARs (97.22\%), and the correlation coefficient of the different $|H_m^c|$ is 0.967. The fitting lines are both close to $y = x$. The spread of the plot symbols is equivalent in the cases of $1''$ and $2''$. These results suggest that the different longitudinal fields (SMFT or MDI) have quite limited contributions to $H_m^c$, and these results also show that the quality of the data preprocessing in Section~\ref{sect:SampSelect} is pretty good for performing the calculation of $H_m^c$.

\subsection{Discussion}

\citet{Yang2013,Yang2018} indicate that the discrepancy between the relative magnetic helicity in the corona and the accumulated helicity injection through the photosphere is less than 10\%. However, these simulations use much smoother magnetic field distributions than the observations, and are without eruptions, and there are still considerable discrepancies (more than 10 times for some samples) between $H_m^p$ ($H_m^p = \int^{T}_{0} \frac{dH}{dt} dt$) and $H_m^c$ ($H_{m,t=T}$) in the observations.

One possible reason is the resolution of the observed magnetograms. Different resolutions not only affect the signs and values of the helicity from the individual methods, but also affect the relation of $H_m^p$ and $H_m^c$. In our estimation, the reduced spatial resolution is more likely to give space for the discrepancy of sign between $H_m^c$ and $H_m^p$, and is more likely to decrease the discrepancy between $|H_m^c|$ and $|H_m^p|$.

The second reason is the net (initial) magnetic helicity before emergence. The net magnetic helicity is not zero, but tends to be consistent with HHR. (The proportion conforming to HHR varies with the resolution and calculation method of $H_m^p$, which is always over 55\%.) However, the reason for the nonzero net magnetic helicity is still unclear.

Third, the evaluations of $H_m^p$ are significantly affected by the different optical flow methods. In our estimation, the $H_m^p$ based on LCT is more sensitive to resolution, and the average discrepancy between the LCT-based $H_m^p$ and DAVE-based $H_m^p$ partly relates to the resolution, which is around 50\% for $1''$ and 7.3\% for $2''$.

Fourth, the calibration coefficients of SMFT are underestimated, especially for the transversal component \citep{Wang2009,Bai2014}. An underestimated vector field $\bm{B}$ leads to an underestimated vector potential $\bm{A}$, and further results in the underestimation of the relative magnetic helicity in the corona $H_m^c$. Considering that the magnetic helicity is proportional to the square of the magnetic field, a slight change of the calibration coefficient will significantly affect the calculation of the magnetic helicity.

Fifth, the magnetic helicity dissipation term is ignored in the formula for magnetic helicity flux (Equation (\ref{equ2})). However, the magnetic Reynolds number is infinite, especially close to the photosphere, so dissipation could affect the process of magnetic helicity transportation, and the turbulent magnetic diffusivity should be taken into account. \citet{Chae2008b} estimated the magnetic diffusivity values at different length scales, finding that they are consistent with a turbulent cascade that ends at a resistive dissipation scale of about 30 km. For full-disk MDI magnetograms with pixel sizes of 1435 km, the mean value of the magnetic diffusivity is about 25 km$^{2}$ s$^{-1}$. In all of our NEARs, the typical transverse velocities of the photospheric fluxes are no more than 0.3 km s$^{-1}$, so the proportion of dissipation terms is around 15\% (thus the estimations of $H_m^p$ are 15\% higher).

In addition, the relative magnetic helicity probably changes because of magnetic helicity exchanging between neighboring NEARs. \citet{Yang2009b} analyzed $H_m^p$ in two NEARs (AR 9188 and AR 9192), finding evidence (the connection of lines) of magnetic helicity exchange between NEARs. No evidence is found by checking the EIT 171\AA\ evolution images corresponding to all of our NEARs, therefore the influence of this point is quite limited. In addition, all 36 of our NEARs are not associated with coronal mass ejections within the time interval of the magnetic helicity injection.

\section{Summary}
\label{sect:Sum}

The estimation and calculation methods for magnetic helicity have been developed over the past few decades \citep{Georgoulis2012,Valori2016,Guo2017}, but the proportion of magnetic helicity transported through the photosphere into the corona is not yet clear. In this paper, we select 36 NEARs in the 23rd solar cycle as a sample, apply optical flow methods to calculate the $H_m^p$ (the accumulated magnetic helicity flux through the photosphere), then adopt the NLFFF model and the finite volume method to estimate the $H_m^c$ (the instantaneous relative magnetic helicity in the corona). The longitudinal magnetograms from SMFT and MDI have been cross-calibrated. Cases from $1''$ and $2''$ magnetograms are estimated to analyze the influence of the resolution. The results show that LCT-based $H_m^p$ is larger than $H_m^c$ in $1''$, and that DAVE-based $H_m^p$ is more consistent with $H_m^c$ than LCT-based $H_m^p$. $H_m^p$ is more consistent with $H_m^c$ in evaluation from $2''$ than from $1''$, however, the deduced spatial resolution is more likely to give space for a discrepancy of sign between $H_m^p$ and $H_m^c$. Moreover, $H_m^c - H_m^p$ systematically shows consistency with HHR (over 55\%), no matter which resolution and method are adopted. The estimation of LCT-based helicity is more sensitive than the other two methods.

We attempt to analyze some of the reasons that are responsible for the discrepancy between $H_m^p$ and $H_m^c$, including the nonzero net (initial) magnetic helicity before emergence, the resolutions of the magnetograms, the different calculation methods, the underestimated calibration coefficients of SMFT, missing the magnetic dissipation term in the formula for magnetic helicity flux, and the magnetic helicity exchange among neighboring regions. The above analyses suggest that the consistency of $H_m^c$ and $H_m^p$ is partly dependent on the resolution of the magnetograms and the calculation methods.

Resolution is important for estimations of magnetic helicity, both in terms of signs and values. In order to research the influence of resolution further, the vector magnetograms from the Helioseismic and Magnetic Imager can be applied at higher resolution ($0.5''$). In addition, these kinds of magnetograms make it possible to estimate the $H_m^c$ successively, so that the errors of cross-calibration from different instruments could be avoided.

\section*{Acknowledgements}
The authors are grateful for the anonymous referees on the detailed comments and suggestions to significantly optimize the manuscript. We also appreciate the HSOS staff and the MDI team providing the excellent magnetograms. This study is supported by the National Natural Science Foundation of China (grants No. 12073040, 11973056, 12003051, 11573037, 12073041, 11427901, and 11611530679); by the Strategic Priority Research Program of the Chinese Academy of Sciences (grants No. XDB09040200, XDA15010700, and XDA15320102); by the Youth Innovation Promotion Association of CAS (2019059); and by the ISSI International Team on Magnetic Helicity estimations in models and observations of the solar magnetic field.


\begin{sidewaystable} \scriptsize \addtolength{\tabcolsep}{-3pt}
\begin{center}
\caption{$H_m^c$ and $H_m^p$ from Different Methods and Different Resolutions}
\label{Tab2}
\renewcommand\arraystretch{1.0}
 \begin{tabular}{ccccccrrrrrrrrr}
  \hline\hline\noalign{\smallskip}
No. & NOAA  & Hemi  & $F_\textrm{\tiny{SMFT}}$\tablenotemark{\scriptsize a} &  $F_\textrm{\tiny{MDI}}$  & \footnotesize{Slope} & $H_\textrm{\tiny{NLFFF,1}}$\tablenotemark{\scriptsize b} & $H_\textrm{\tiny{NLFFF,2}}$ & $H_\textrm{\tiny{COM,1}}$ & $H_\textrm{\tiny{COM,2}}$ & $H_\textrm{\tiny{LCT,1}}$ & $H_\textrm{\tiny{LCT,2}}$ & $H_\textrm{\tiny{DAVE,1}}$ & $H_\textrm{\tiny{DAVE,2}}$ & $H_\textrm{\tiny{NLFFF,2}}-H_\textrm{\tiny{LCT,2}}$ \\
  \noalign{\smallskip}\hline\noalign{\smallskip}
1  & 7988  & S & 10.546 & 14.464 & 0.97 & 0.6801    & 0.7035    & 0.8538    & 0.7024    & $-0.5699$ & $-1.5624$ & $-1.5417$ & $-2.0954$ & 2.2659    \\
2  & 8026  & S & 72.347 & 65.441 & 0.86 & $-31.344$ & $-54.966$ & $-21.389$ & $-43.022$ & $-29.983$ & $-90.074$ & $-7.7800$ & $-38.030$ & 35.108    \\
3  & 8032  & S & 27.943 & 39.497 & 1.16 & 24.074    & 38.948    & 23.945    & 47.653    & 34.785    & 30.764    & 11.323    & 17.942    & 8.1839    \\
4  & 8052  & N & 19.826 & 35.048 & 1.43 & 4.7167    & 8.1977    & 7.1640    & 9.2655    & 23.022    & 22.251    & 11.202    & 16.417    & $-14.053$ \\
5  & 8093  & N & 25.064 & 65.221 & 2.34 & $-4.6639$ & $-19.412$ & 1.2591    & $-9.3317$ & 90.275    & 55.799    & 57.939    & 92.771    & $-75.212$ \\
6  & 8122  & N & 27.111 & 62.000 & 2.08 & 11.880    & 18.820    & 4.7300    & 8.4082    & 117.49    & 84.383    & 30.934    & 28.729    & $-65.563$ \\
7  & 8123  & N & 31.757 & 63.045 & 2.22 & 89.249    & 137.91    & 53.480    & 103.92    & 235.37    & 196.80    & 201.53    & 258.87    & $-58.896$ \\
8  & 8164  & N & 31.301 & 80.748 & 2.45 & $-134.64$ & $-236.47$ & $-127.32$ & $-204.16$ & $-297.46$ & $-166.98$ & $-104.40$ & $-171.17$ & $-69.497$ \\
9  & 8205  & N & 24.492 & 48.781 & 1.64 & $-14.662$ & $-25.880$ & $-13.56$  & $-25.880$ & $-26.848$ & $-7.6521$ & $-25.771$ & $-39.970$ & $-18.228$ \\
10 & 8226  & N & 49.288 & 82.416 & 1.54 & $-13.980$ & $-32.700$ & 1.4000    & 21.460    & $-225.88$ & $-87.065$ & $-88.944$ & $-91.053$ & 54.365    \\
11 & 8404  & S & 63.111 & 111.06 & 1.78 & $-316.35$ & $-404.75$ & $-290.13$ & $-383.29$ & $-315.59$ & $-276.15$ & $-144.62$ & $-218.28$ & $-128.60$ \\
12 & 8582  & N & 87.382 & 110.58 & 1.16 & $-264.86$ & $-376.75$ & $-255.24$ & $-366.15$ & $-335.18$ & $-213.60$ & $-179.27$ & $-223.44$ & $-163.15$ \\
13 & 8722  & N & 51.696 & 73.482 & 1.33 & 1.0508    & $-1.9584$ & $-1.5764$ & $-12.163$ & 91.241    & 15.976    & 126.23    & 156.47    & $-17.934$ \\
14 & 9058  & S & 23.021 & 42.520 & 1.52 & $-14.199$ & $-19.371$ & $-11.669$ & $-20.561$ & $-29.731$ & $-25.237$ & $-7.0931$ & $-5.1213$ & 5.8658    \\
15 & 9170  & S & 72.267 & 82.261 & 1.00 & 51.573    & 89.848    & 47.744    & 108.30    & 284.07    & 71.979    & 127.04    & 153.02    & 17.868    \\
16 & 9297  & N & 22.121 & 41.300 & 1.44 & 7.9179    & 14.314    & $-7.8234$ & 9.0389    & $-0.8762$ & $-8.3361$ & 1.1403    & $-7.7049$ & 22.650    \\
17 & 9366  & S & 62.819 & 76.688 & 1.02 & $-29.674$ & $-114.28$ & $-28.189$ & $-88.737$ & 76.328    & 82.666    & 85.825    & 120.11    & $-196.94$ \\
18 & 9396  & S & 33.984 & 77.843 & 2.10 & 37.551    & 71.199    & 41.936    & 79.599    & 6.2934    & $-72.712$ & 12.379    & 12.459    & 143.91    \\
19 & 9417  & S & 49.922 & 84.479 & 1.43 & 59.832    & 95.935    & 91.294    &  143.62   & 101.79    & 60.698    & 52.902    & 70.416    & 35.238    \\
20 & 9456  & N & 39.185 & 59.728 & 1.34 & 7.3409    & 17.451    & 12.404    & 24.099    & 73.726    & 44.390    & 35.098    & 52.845    & $-26.939$ \\
21 & 9533  & S & 28.617 & 50.772 & 1.50 & 11.714    & 19.366    & 18.290    & 28.872    & 37.126    & 14.193    & $-5.1202$ & $-11.304$ & 5.1736    \\
22 & 9559  & S & 15.446 & 28.363 & 1.46 & 9.3220    & 15.502    & 12.667    & 24.722    & 24.092    & 11.321    & 17.029    & 24.869    & 4.1815    \\
23 & 9692  & N & 37.911 & 143.15 & 3.44 & $-110.37$ & $-160.04$ & $-165.97$ & $-272.37$ & $-473.54$ & $-397.03$ & $-138.14$ & $-202.46$ & 236.99    \\
24 & 9873  & S & 28.817 & 96.433 & 3.84 & 149.74    & 221.27    & 120.24    & 216.10    & $-471.96$ & $-213.36$ & $-180.26$ & $-281.71$ & 434.63    \\
25 & 9924  & S & 27.970 & 58.307 & 2.36 & $-14.350$ & $-40.072$ & $-19.866$ & $-41.198$ & $-2.0703$ & $-17.749$ & $-4.7059$ & $-3.6100$ & $-22.323$ \\
26 & 10045 & S & 23.354 & 64.356 & 2.84 & 39.730    & 68.336    & 39.949    & 74.439    & $-8.1930$ & $-14.775$ & 16.129    & 29.799    & 83.112    \\
27 & 10057 & S & 34.740 & 79.907 & 2.08 & 49.966    & 84.139    & 66.290    & 105.85    & 77.670    & 26.257    & 42.224    & 62.584    & 57.881    \\
28 & 10132 & N & 31.011 & 94.837 & 2.84 & 220.08    & 322.44    & 263.86    & 366.37    & 651.38    & 474.31    & 342.68    & 504.95    & $-151.87$ \\
29 & 10188 & N & 26.525 & 60.957 & 2.50 & $-73.620$ & $-120.83$ & $-76.440$ & $-122.58$ & $-139.68$ & $-25.352$ & $-77.440$ & $-87.480$ & $-95.478$ \\
30 & 10268 & N & 14.630 & 48.329 & 4.04 & $-118.82$ & $-176.80$ & $-106.16$ & $-142.53$ & $-7.9116$ & 3.2553    & $-15.430$ & $-23.874$ & $-180.06$ \\
31 & 10692 & S & 17.378 & 39.356 & 2.14 & 2.4605    & 9.8629    & 4.7311    & 14.543    & $-112.01$ & $-90.393$ & 7.7930    & 45.204    & 100.26    \\
32 & 10747 & S & 13.808 & 47.490 & 3.59 & 10.891    & 12.995    & 12.649    & 12.851    & $-2.0447$ & 3.3002    & $-15.123$ & $-25.226$ & 9.6951    \\
33 & 10813 & S & 16.901 & 67.508 & 5.11 & $-67.947$ & $-96.654$ & $-74.954$ & $-126.51$ & $-212.65$ & $-178.02$ & $-127.63$ & $-149.79$ & 81.366    \\
34 & 10838 & N & 25.785 & 69.802 & 3.61 & $-40.634$ & $-69.189$ & $-10.672$ & $-23.796$ & $-222.56$ & $-63.706$ & $-65.846$ & $-68.241$ & $-5.4829$ \\
35 & 10848 & S & 30.685 & 121.26 & 4.36 & 163.65    & 226.68    & 112.31    & 198.39    & $-171.77$ & $-80.463$ & 137.77    & 181.66    & 307.14    \\
36 & 10869 & S & 13.138 & 37.914 & 2.90 & $-7.8069$ & $-20.576$ & $-9.7788$ & $-29.159$ & 8.2820    & $-4.0866$ & 3.3761    & 4.6432    & $-16.490$ \\
  \noalign{\smallskip}\hline
\end{tabular}
\end{center}
\tablenotetext{}{\scriptsize \textbf{Notes.} $F_\textrm{\tiny{SMFT}}$ and $F_\textrm{\tiny{MDI}}$ are the longitudinal magnetic fluxes, columns 7--10 are the instantaneous relative magnetic helicity in the corona, and columns 11--14 are the accumulated magnetic helicity flux through the photosphere. The subscripts ``1'' and ``2'' denote the resolutions of $1''$ and $2''$, respectively.}
\tablenotetext{a}{\ \scriptsize The unit of magnetic flux is $10^{20}$ Mx.}
\tablenotetext{b}{\ \scriptsize The unit of $H_m^p$ and $H_m^c$ is $10^{40}$ Mx$^2$. The sign $-$ ($+$) in the helicity column represents negative (positive) helicity.}
\end{sidewaystable}

\end{document}